\documentstyle[multicol,aps,psfig]{revtex}
\begin{document}

\newcommand{\lsim}
{\ \raisebox{2.75pt}{$<$}\hspace{-8.0pt}\raisebox{-2.75pt}{$\sim$}\ }
\newcommand{\gsim}
{\ \raisebox{2.75pt}{$>$}\hspace{-8.0pt}\raisebox{-2.75pt}{$\sim$}\ }

\title{Why the observed jet quenching at RHIC is due to parton energy loss}
\author{Xin-Nian Wang}
\address{
Nuclear Science Division, MS70R0319,\\
Lawrence Berkeley National Laboratory, Berkeley, CA 94720}

\date{April 20, 2003}
\maketitle

\vspace{-1.5in}
{\hfill LBNL-53275}
\vspace{1.4in}

\begin{abstract}
Significant jet quenching in central $Au+Au$ collisions
has been discovered at RHIC. This paper provides theoretical arguments 
and lists experimental evidence that the observed
jet quenching at RHIC is due to parton energy loss instead of
hadron rescattering or absorption in a hadronic medium.
These include: (1) hadron formation time based on 
the uncertainty principle, (2) $p_T$ dependence and (3) centrality dependence 
of the observed jet quenching, (4) jet-like leading hadron correlations 
(5) high-$p_T$ azimuthal anisotropy and (6) experimental 
data from $Pb+Pb$ collisions at SPS and $e+A$ collisions.
Direct measurements of the parton energy loss in the direction of 
a triggered high-$p_T$ hadron and the medium modified fragmentation 
function on the back-side are proposed to further verify the 
partonic nature of the observed jet quenching. The importance of 
jet quenching studies at lower energies at RHIC is also discussed.

\end{abstract}

\begin{multicols}{2}

\section{Introduction}

More than a decade after the original proposal \cite{gp90,wg92}
of jet quenching due to radiative parton energy loss, conclusive 
experimental evidence has been found in central $Au+Au$ collisions 
at the Relativistic Heavy-ion Collider (RHIC) not only from the 
suppression of high-$p_T$ single inclusive hadron
spectra \cite{phenix-r,star-r,phobos-r} but also the suppression
of back-side jet-like correlations \cite{star-c}. The latter 
provides direct evidence for medium modification of the parton
fragmentation  functions \cite{whs}. More recent results of $d+Au$ 
collisions \cite{phenixda,starda,phobosda} further
prove that the observed jet quenching is due to final-state
interactions with the produced medium. Initial-state scatterings
in cold nuclei only broaden the initial transverse momentum, leading
to the Cronin enhancement of intermediate high-$p_T$ hadron 
spectra as was first predicted for $p+A$ collisions at RHIC \cite{wang98}.

The original proposal of jet quenching in a dense (or normal) nuclear
medium \cite{gp90,wg92} was based on the idea that radiative energy
loss during the propagation of an energetic parton must suppress
the leading hadron distributions inside a jet. This leads to
medium modification of the jet fragmentation functions \cite{whs}
and suppression of the high-$p_T$ hadron spectra in 
high-energy heavy-ion collisions. 
Such medium-induced radiative parton energy loss has since been 
studied in detail and in many different 
approaches \cite{gw93,bdmps,zak,glv00,wied00,guow,ww02} 
in QCD that include the non-Abelian Laudau-Pomerachuck-Migdal (LPM) 
interference effect. The energy loss was found to be proportional
to the gluon density of the medium. It was further predicted that 
jet quenching due to parton energy loss should also lead to the 
azimuthal anisotropy of high-$p_T$ hadron spectra in non-central 
heavy-ion collisions \cite{wangv2}, which has been 
observed \cite{star-v2} at RHIC.

Phenomenological studies of hadron 
spectra based on parton energy loss have found that the observed
suppression of high-$p_T$ single hadron spectra implies large parton
energy loss or high initial gluon 
density \cite{ww02,wangqm02,jeon,mueller,gv02}. The same
parton energy loss is also found to reproduce the observed
suppression of back-side correlation \cite{hirano,wang03} and
the high-$p_T$ azimuthal anisotropy \cite{wang03,gvw}. Most
importantly, the calculated centrality dependences of the suppression of both 
single hadron spectra and back-side correlation agree very
well with the experimental measurements \cite{wang03}. The
deduced initial gluon density at an initial time $\tau_0=0.2$ fm/$c$
is found to be about 30 times of that in a normal nuclear 
matter \cite{gv02,wang03}. If the transverse energy per particle is
0.5 GeV \cite{phenix-et}, the above gluon density will correspond to
an initial energy density of $\epsilon=15$ GeV/fm$^3$, which is
about 100 times of the energy density in a cold nuclear matter.
In addition, the measured large 
azimuthal anisotropy for soft hadrons is found to saturate 
the hydrodynamic limit \cite{starv2-2,phenix-v2}.
These experimental results all point to an initial
medium that is strongly interacting and has a large initial 
pressure gradient. Within our current understanding of QCD, such a 
strongly interacting medium with about 100 times normal nuclear 
energy density can no longer be a normal hadronic matter.

The aforementioned analyses of RHIC data on jet quenching are all
based on a picture in which partons propagating through the dense
medium lose energy first and then hadronize outside in the same way as 
in the vacuum. It is reasonable to ask whether leading hadrons 
from the jet fragmentation could have strong interaction with 
the medium and whether hadron absorption could be the main cause 
for the observed jet quenching. 
This paper will provide arguments against such a scenario in detail and 
list experimental evidence that the observed patterns of jet quenching 
in heavy-ion collisions at RHIC can {\it only} be the consequences of parton 
energy loss, not hadronic absorption.

\section{Hadron formation time}

Fragmentation of a parton into hadrons involves mainly non-perturbative
physics in QCD and thus is not calculable within perturbative QCD (pQCD). 
One can, nevertheless, use pQCD to calculate the evolution of the 
fragmentation process due to short distance interaction when the 
virtuality of the parton is larger than $Q_0\sim 1$ GeV.
Such perturbative processes can take place over a period of time,
\begin{equation}
\tau_{\rm DGLAP}\sim 2\sum_i \frac{Ez_i(1-z_i)}{Q^2_i} 
\gsim 2\frac{Ez_0(1-z_0)}{Q^2_0}\; ,
\end{equation}
where the sum is over gluon emission, and $Q_i$ and $z_i$ 
are the virtualities and fractional energies of the
intermediate partons between each successive emission 
until $Q_0$ is reached. Afterwards, the non-perturbative processes 
of hadronization take place. One scenario of the non-perturbative
process is that the produced partons (quarks and gluons) will
recombine into the final hadrons. The hadron formation time 
can be considered as the time for partons to 
build up their color fields and develop the hadron wave function. 
According to the uncertainty principle, such a formation time in the
rest frame of the hadron can be related to the hadron size $R_h$. In
the laboratory frame, the hadron formation time is then \cite{dok}
\begin{equation}
\tau_f\sim R_h\frac{E_h}{m_h} \; .
\end{equation}
For an $E_h=10$ GeV pion, this amounts to $\tau_f\sim 35-70$ fm/$c$ for
$R_h=0.5-1$ fm.

In some dipole models of hadronization \cite{kopel}, the quarks and
anti-quarks from gluon splitting are assumed to combine into color
singlet dipoles which will become the final hadrons. The hadron 
formation time is then assumed to be just the formation time for
the gluon emission, ignoring the time of quark and anti-quark 
production and the time for dipoles to grow to the normal hadron size. 
Even if one considers this alternative hadronization process as 
successive emission of hadrons by the fragmenting jet,
a hadron carrying a fraction $z$ of the parton energy will take
\begin{equation}
\tau_f\sim \frac{2E_h(1-z)}{k_T^2+m_h^2}
\end{equation}
to be produced, where $k_T\sim \Lambda_{\rm QCD}$ is the intrinsic
transverse momentum of the hadron. 
As we will show later, a 10 GeV hadron comes from a parton
with an average energy $E=16.5$ GeV in $p+p$ collisions at RHIC,
thus an average $\langle z\rangle =0.6$.
Using $\Lambda_{\rm QCD}=0.2$ GeV, the formation time for a 10 GeV pion
is then $\tau_f\sim 40$ fm/$c$.

Though the above numbers can only serve as order-of-magnitude
estimates, they are still much longer than the typical medium size 
or the lifetime of
the dense medium in heavy-ion collisions at RHIC. Furthermore, the
above estimates are for hadronization in vacuum only. Medium 
interaction with the fragmenting partons will only increase
the hadron formation time. Certainly, in the extreme case, the 
hadron can never be formed inside a deconfined medium due to
color screening and the formation time should 
never be shorter than the lifetime of a quark-gluon plasma.

\section{Momentum dependence of hadron suppression}

The most striking feature of the observed jet quenching manifested 
in the suppression of high-$p_T$ hadrons is the almost flat $p_T$ 
dependence of the suppression at high $p_T$ \cite{phenix-r,star-r}.
The empirical total energy loss has to have a linear energy
dependence in order to describe such a $p_T$ 
dependence \cite{jeon,mueller}. This runs directly opposite to 
the trend of hadronic absorption or rescattering. Since the 
hadron formation time is proportional to the hadron or jet energy, 
the total effective energy loss due to hadron rescattering or 
absorption should decrease with energy, unless the energy 
dependence of the hadronic energy loss per unit distance is
stronger than a quadratic dependence. Such a quadratic or stronger
energy dependence of the energy loss can never be
allowed in any physical scenario.

For elastic scatterings, the energy loss of a pion per scattering
is $\Delta E_{\rm el}\approx E_{\pi}(1-\cos\theta_{\rm cm})/2$, where
$\theta_{\rm cm}$ is the scattering angle in the center of mass frame.
The averaged elastic energy loss can be estimated as
\begin{equation}
\frac{dE_{\rm el}}{dx}
=\langle \int dt\frac{d\sigma}{dt}E_{\pi}\frac{-t}{s}\rho_h\rangle
\approx \frac{\sigma_0}{B} \langle \frac{\rho_h}{\omega_h}\rangle\; ,
\end{equation}
which has a very weak energy dependence. 
Here $t\approx -s(1-\cos\theta_{\rm cm})/2$, $s\approx 2E_{\pi}\omega_h$
and $\langle\cdots\rangle$ is the thermal average over 
hadron energy $\omega_h$ with density $\rho_h(\omega_h)$. 
We have considered only the dominant $t$-channel
when $\sqrt{s}$ is much larger than the $\pi$-$h$ resonance mass 
and $d\sigma/dt$ can be described by its geometrical 
form $d\sigma/dt=(\sigma_0 B)\exp(tB)$,
with $B/\sigma_0\approx$ 0.3 according to the observed 
geometrical scaling property of high energy hadron 
collisions for $\sqrt{s}<100$ GeV \cite{gscale}.
Here, $\sigma_0$ is assumed to be
the  total cross section. Normally, elastic cross section is 
about 17\% of the total cross section.
This elastic energy loss is also related to the transverse momentum 
broadening,
\begin{equation}
\langle \frac{q_T^2}{\lambda}\rangle
\approx \frac{\sigma_0}{B} \langle\rho_h\rangle \; .
\end{equation}
For a pion gas at $T\sim 150$ MeV, the elastic energy loss
is very small, about 0.036 GeV/fm, independent of the 
pion's energy. The corresponding transverse momentum 
broadening will be also very small.
The energy loss due to inelastic $\pi$-$h$ scattering is difficult
to estimate. However, it should not have a linear energy 
dependence, according to the estimate based on the uncertainty 
principle \cite{brodsky}, taking into account the LPM interference
effect. Therefore, the energy loss due to hadronic interaction
should have an energy dependence weaker than a linear dependence.
Hadronic rescattering or absorption, with the energy dependence 
of the formation time, cannot give rise to the
observed flat $p_T$ dependence of the hadron suppression.

\begin{figure}
\centerline{\psfig{figure=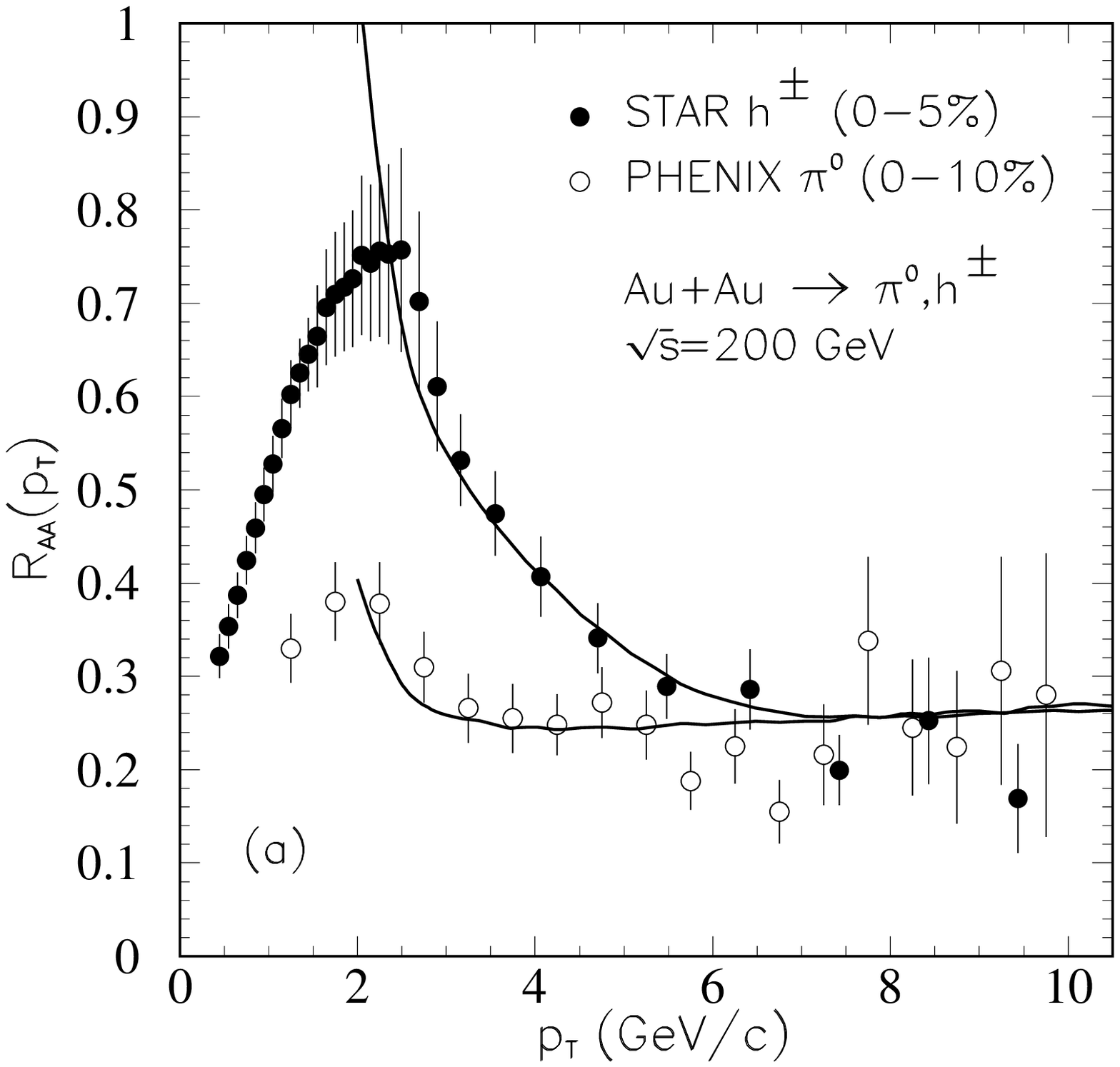,width=3.0in,height=2.5in}}
\centerline{\psfig{figure=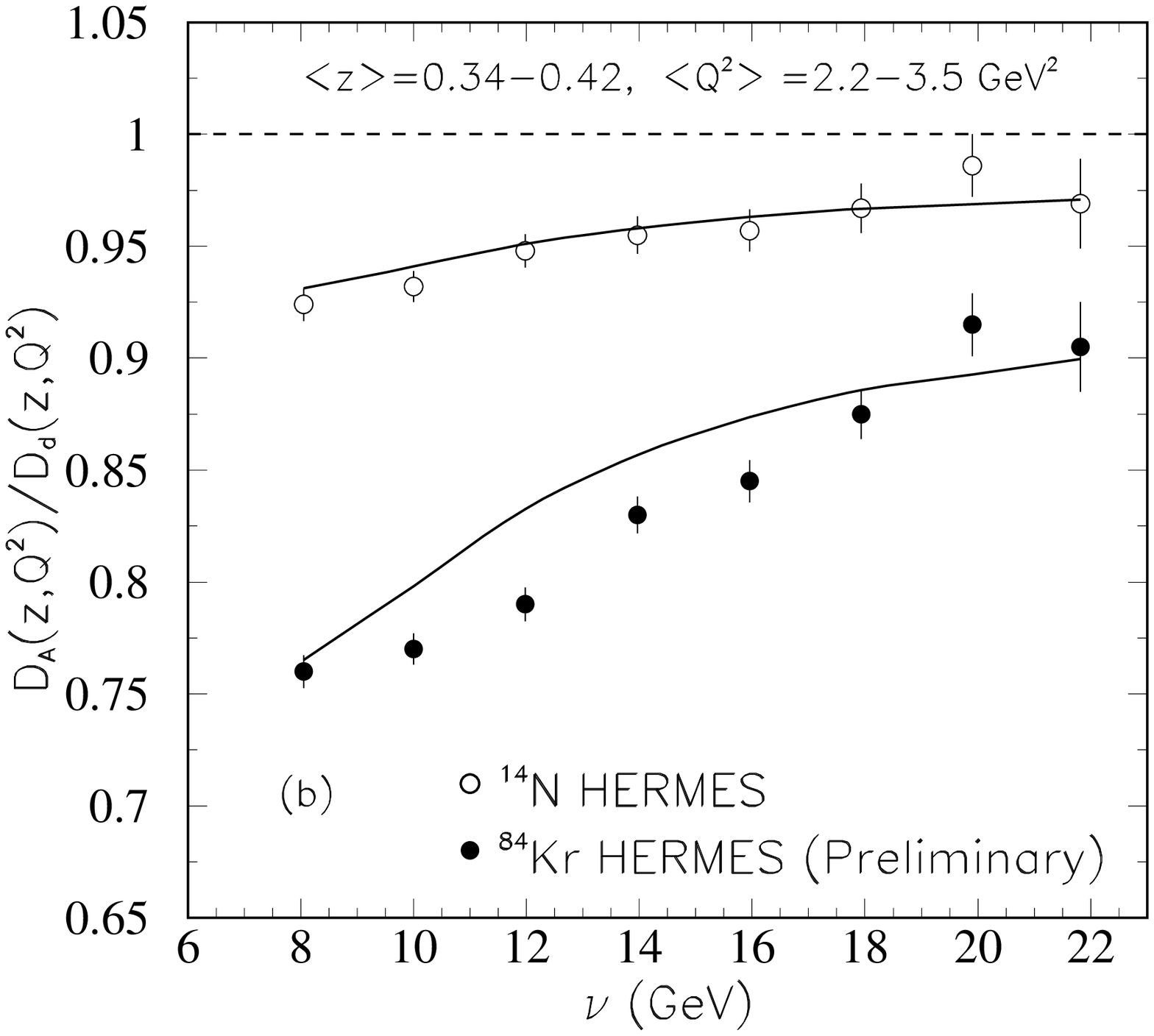,width=3.0in,height=2.5in}}
\caption{The suppression factor for (a) single inclusive hadron
spectra in central $Au+Au$ collisions and (b) deeply inelastic
scattering off nuclear targets. Solid lines are theoretical
calculations with parton energy loss and modified fragmentation
functions. Data are from PHENIX  \protect\cite{phenix-r}, 
STAR \protect\cite{star-r} and HERMES  \protect\cite{hermes}.}
\label{fig1}
\end{figure}

If hadronic rescattering or absorption were responsible for 
jet quenching, it most likely would happen in deeply inelastic
$eA$ scattering (DIS), where a quark jet propagates through
the normal nuclear matter before its hadrons are detected.
As shown in Fig.~\ref{fig1}(b), the suppression of leading hadrons
clearly disappears as the initial quark energy is increased. 
The calculation of modified parton fragmentation functions
due to parton rescattering and gluon bremsstrahlung \cite{guow,ww02,arleo},
as shown by the solid lines, agrees with
the data very well, though the data can also be 
explained \cite{cassing} as a consequence of the 
hadron absorption after a short formation time.
The hadron suppression in the central $Au+Au$ collisions at RHIC, 
on the other hand, are almost constant at high $p_T$ as shown in 
Fig.~\ref{fig1}(a). This is hard to understand from
the original theoretical calculations of parton energy loss,
since the results only depend on the gluon density, whether in a cold 
or hot medium. The different energy dependences in DIS and heavy-ion
collisions can be explained by the effect of absorption of thermal 
gluons from a thermal bath, which only exists in heavy-ion 
collisions but is absent in DIS. This detailed
balance between gluon emission and absorption in a hot medium increases
the energy dependence of the net energy loss \cite{ww01}. The
solid lines in Fig.~\ref{fig1}(a) are calculations based on a
parameterization of the energy loss that includes the effect
of detailed balance. Calculations shown in Fig.~\ref{fig1}(b) as solid
lines for DIS only include induced gluon radiation in cold nuclei.

\section{Centrality dependence of hadron suppression}

This paper will not describe the details of the calculation of single hadron 
and dihadron spectra in heavy-ion collisions, but refer readers to
Ref.\cite{wang03}. It is, however, important to point out that the 
effective total parton energy loss in a dynamic system is
proportional to a path integral of the gluon density
along the propagation trajectory.
According to recent theoretical studies \cite{ww02,gvw,sw02},
\begin{equation}
\Delta E\approx \langle \frac{dE}{dL}\rangle_{1d}
\int_{\tau_0}^{\tau_0+\Delta L} d\tau \frac{\tau-\tau_0}{\tau_0\rho_0}
\rho_g(\tau,b,\vec{r}+\vec{n}\tau), \label{eq:de}
\end{equation}
where $\rho_0$ is the averaged initial gluon density at $\tau_0$ 
in a central collision, and $\langle dE/dL\rangle_{1d}$ 
is the average parton energy loss over a distance $R_A$
in a 1-dimensional expanding medium with an initial uniform gluon 
density $\rho_0$. The corresponding energy loss 
in a static medium with a uniform gluon density 
$\rho_0$ over a distance $R_A$ is 
$dE_0/dL=(R_A/2\tau_0)\langle dE/dL\rangle_{1d}$ \cite{ww02}.
The gluon density $\rho_g(\tau_0,r)$ is assumed to be 
proportional to the transverse profile of participant nucleons,
which is consistent up to 30\% with the measured charged 
hadron multiplicity \cite{phenixn,phobosn}.

\begin{figure}
\centerline{\psfig{figure=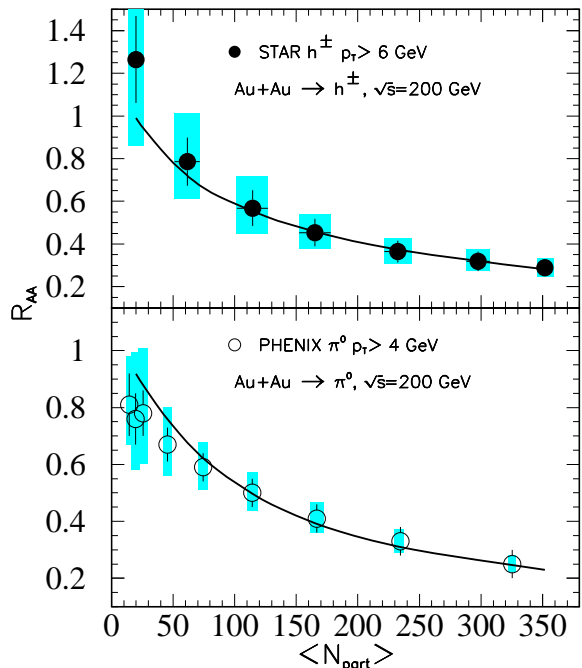,width=3.0in,height=3.5in}}
\caption{The centrality dependence of the measured single inclusive 
hadron suppression \protect\cite{phenix-r,star-r} at high $p_T$ as 
compared to theoretical calculation with parton energy loss}
\label{fig2}
\end{figure}

The calculated centrality dependence of the single hadron
suppression in $Au+Au$ collisions agrees very well with the
experimental measurements, as shown in Fig.~\ref{fig2}. The
centrality dependence of the back-side suppression is also in
excellent agreement with the data \cite{wang03}. These
are the consequences of the centrality dependence of the
averaged total energy loss in Eq.~(\ref{eq:de}) and the
surface emission of the surviving jets. Jets produced around
the core of the overlapped region are strongly suppressed,
since they lose the largest amount of energy.

On the other hand, if the finite hadron formation time were 
shorter than the medium size in the most central collisions and
jet quenching were only caused by the subsequent rescattering or 
absorption of the leading hadrons, one should expect a more
rapid disappearance or reduction of jet quenching when the medium 
size becomes smaller than the hadron formation time in non-central
$Au+Au$ collisions. This is clearly absent in the observed 
centrality dependence.

The large suppression of single hadron spectra, about a factor of 5,
in the most central $Au+Au$ collisions can actually lead to 
a strong constraint on the
hadron formation time if no parton energy loss is allowed.
One can take the most extreme scenario: There is no jet
attenuation before a finite hadron formation time $\tau_f$ and
every hadron is absorbed if it is still inside the medium at the
formation time. The suppression factor is then determined by the
ratio of the volume of the outer layer with a width $\tau_f$ and
the total overlapping volume. Here one neglects the finite transverse
flow velocity in the early time. With a hard-sphere nuclear 
geometry, one finds that a factor of 5 suppression would require
a formation time shorter than 2 fm/$c$, which is hard to 
reconcile with the theoretical estimate for a 10 GeV pion.

\section{Jet-like hadron correlation}

Perhaps the most discriminating experimental evidence against 
jet quenching via hadron rescattering or absorption comes from
two-particle correlations. Jet structure of azimuthal
correlations of leading hadrons is clearly seen in RHIC 
experiments and it is the same in $p+p$, $d+Au$ and 
peripheral $Au+Au$ collisions \cite{star-c,starda}.
It consists of one peak in the near-side of the triggered hadron and 
another in the back-side. As one increases the centrality
in $Au+Au$ collisions, the back-side correlation is significantly
suppressed just as the single hadron spectra. The 
near-side correlation, on the other hand, remains the same as
in $p+p$ and $d+Au$ collisions. This is clear evidence that jet
hadronization takes place outside the dense medium with a reduced parton
energy. On the other hand, let us suppose that the leading and 
sub-leading hadrons from jet fragmentation are produced inside 
the dense medium, hadron rescattering and absorption will certainly 
change the near-side correlation as a function of centrality 
both in strength and shape,
if they are responsible for the suppression 
of single hadron spectra and back-to-back correlations.
Barring corrections due to trigger bias toward
surface emission, the same-side correlation should be suppressed
as much as the single hadron spectra, were the suppression caused
by hadron absorption. This is clearly not seen in the data.
This measurement of leading and subleading hadron correlation
can also be employed in the DIS experiment to study hadron
formation. When the initial quark energy is sufficiently
small, hadrons will be formed inside the nucleus and the
jet profile will be modified due to hadronic rescattering
or absorption.

It should be stressed that the above argument is only true when
the transverse momenta of the leading and subleading hadrons
are close to each other. This is to ensure that both of them 
come from hadronization of the leading parton. If the subleading
hadron is very soft, then contribution from emitted gluons
induced by bremsstrahlung can become important. These soft
hadrons will then have different correlation and azimuthal
profile from that in $pp$ collisions.

\begin{figure}
\centerline{\psfig{figure=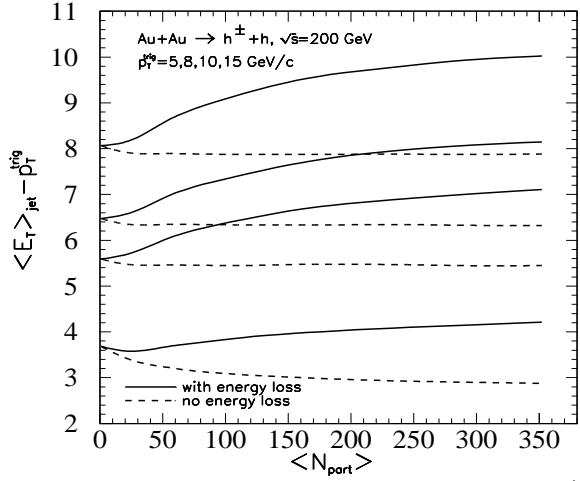,width=3.0in,height=2.5in}}
\caption{The average transverse 
energy $\langle E_T\rangle^{jet}-p_T^{\rm trig}$
of the initial partons that produce a final hadron with $p_T^{\rm trig}$
as a function of $\langle N_{\rm part}\rangle$ for different values
of $p_T^{\rm trig}$ (increasing from lower to top lines). 
Solid lines are for $Au+Au$ collisions with finite parton energy loss 
that describes the inclusive hadron suppression and dashed lines for
calculation without parton energy loss (but with initial multiple
scatterings).}
\label{fig3}
\end{figure}

Because of the trigger bias, the triggered high-$p_T$ hadrons 
mainly come from jets that are produced near the surface 
of the overlapped region. However, on the average the original jet 
should lose a finite amount of energy. In the pQCD parton model, one can
calculate the average energy of the initial jet that, after 
rescattering and induced bremsstrahlung, eventually produces a 
leading hadron with transverse momentum $p_T^{\rm trig}$. Shown
in Fig.~\ref{fig3} are the averaged jet energies minus $p_T^{\rm trig}$
as functions of $\langle N_{\rm part}\rangle$ for different 
values of $p_T^{\rm trig}$. Note that the averaged
$\langle z\rangle=p_T^{\rm trig}/\langle E_T\rangle^{jet}$ in
$p+p$ collisions is about 0.6-0.7, with the triggered hadron 
carrying most of the jet energy. In heavy-ion collisions,
the jet loses some amount of energy before it hadronizes. Therefore,
it has to have higher initial energy than in $p+p$ collisions
in order to produce a leading hadron with the same $p_T^{\rm trig}$.
The extra amount of energy increases with centrality as shown by
the solid lines. 

\begin{figure}
\centerline{\psfig{figure=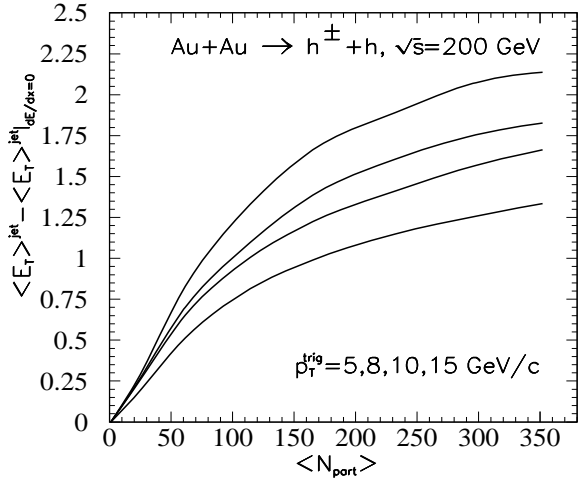,width=3.0in,height=2.5in}}
\caption{The average energy loss for partons that produce a final
hadron with $p_T^{\rm trig}$ in $Au+Au$ collisions.}
\label{fig4}
\end{figure}

Note that $\langle E_T\rangle^{jet}$ evaluated here is the
transverse energy in the center of mass frame of the two colliding 
partons. Initial multiple scattering will increase the initial 
parton transverse momentum leading to the observed Cronin enhancement 
of high-$p_T$ single hadron spectra in $d+Au$ 
collisions \cite{wang98,starda,phenixda,phobosda}.
The trigger bias then leads to smaller values 
of $\langle E_T\rangle^{jet}$ in $Au+Au$ collisions without energy 
loss than in $p+p$ for a fixed $p_T^{\rm trig}$ as shown by the dashed
lines in Fig.~\ref{fig3}. The difference between solid and dashed
lines should then be the averaged energy loss for a jet that survived
multiple scattering and gluon bremsstrahlung and produces a leading
particle with $p_T^{\rm trig}$. This is shown in Fig.~\ref{fig4}
as a function of $\langle N_{\rm part}\rangle$. In the most central
collisions, jets that produce a leading hadron 
at $p_T^{\rm trig}=5-15$ GeV/$c$ lose about $1.4-2.2$ GeV energy on
the average.

In experimental determination of the initial jet energy, one has
to count all hadrons in the jet cone, including those coming from
the emitted gluons. Some of these hadrons are very soft. One
has to use a momentum cut-off as small as possible in order to
make sure the measured total energy is as close to the true value
of the initial jet energy as possible. Since soft hadrons from
medium induced gluons could have a broader angular distribution,
one should have a large jet cone with $|\Delta\phi|<\pi/2$.

\begin{figure}
\centerline{\psfig{figure=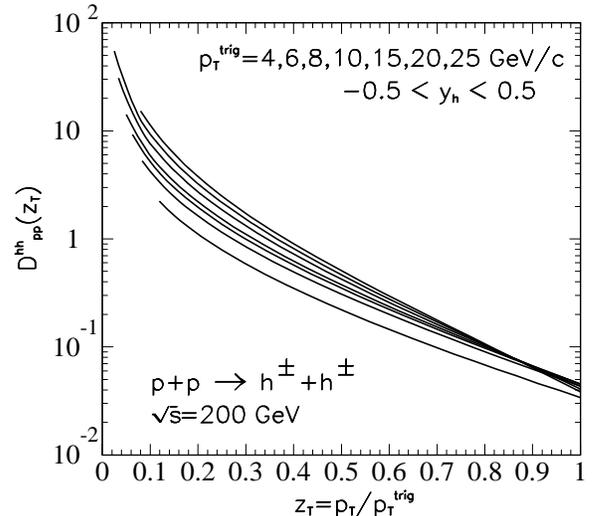,width=3.0in,height=2.7in}}
\caption{Hadron-triggered effective fragmentation function in
$p+p$ collisions from pQCD parton model calculation for different
values of $p_T^{\rm trig}$ (with increasing values
from lower to top solid lines).}
\label{fig5}
\end{figure}

On the back-side of the triggered hadrons, one can define a 
hadron-triggered effective fragmentation function \cite{wang03},
\begin{equation}
  D^{h_1h_2}(z_T,p^{\rm trig}_T)=
  p^{\rm trig}_T \frac{d\sigma^{h_1h_2}_{AA}/dp^{\rm trig}_T dp_T}
  {d\sigma^{h_1}_{AA}/dp^{\rm trig}_T},
\end{equation}
for associated hadron $h_2$ with $p_T$ in the back-side direction 
of $h_1$ with $p_T^{\rm trig}$, where $z_T=p_T/p_T^{\rm trig}$.
The back-side direction is defined by $|\Delta\phi-\pi|<\pi/2$.
This way, one can ensure that the jet cone includes most of the soft
hadrons. This is equivalent to finding remnants of lost jets
in heavy-ion collisions \cite{pratt}.
Shown in Fig.~\ref{fig5} are the hadron-triggered fragmentation
functions in $pp$ collisions.
The differences between different values of $p_T^{\rm trig}$
are caused by scale dependence of the parton fragmentation
functions and the different parton flavor composition, in
particular the ratio of quark and gluon jets. The parton
fragmentation functions used in the calculation are
given by parameterization. With finite values of initial jet
energy, mass and other higher-twist corrections become important.
The actual fragmentation functions will saturate and decrease for
small values of $z_T$. The larger the $E_T$, the smaller the $z_T$
of the saturation point.

\begin{figure}
\centerline{\psfig{figure=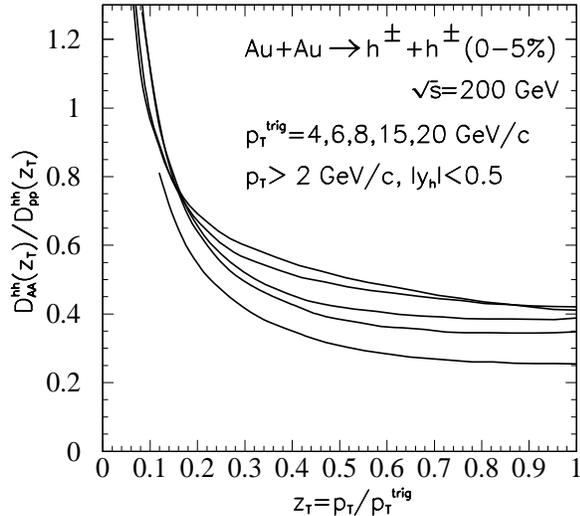,width=3.0in,height=2.7in}}
\caption{The medium modification of the hadron-triggered fragmentation 
function, defined as the ratio of hadron-triggered fragmentation functions
in central $Au+Au$ and $p+p$ collisions for different values 
of $p_T^{\rm trig}$ (increasing from lower to top solid lines).}
\label{fig6}
\end{figure}

Shown in Fig.~\ref{fig6} are the ratios of the hadron-triggered
fragmentation functions in central $Au+Au$ and $p+p$ collisions.
In $Au+Au$ collisions, hadrons are produced not only from
jet fragmentation of the leading partons with reduced energy 
but also from the hadronization of the medium induced gluons.
Normally, hadrons from medium induced gluons are softer and have
a wider angular distribution than the hadrons from leading partons.
One therefore has to define a bigger jet cone for the
hadron-triggered effective fragmentation function.
The softening of the effective fragmentation function is caused
by the suppression of leading hadrons due to parton energy loss
and the enhancement of soft hadrons from emitted gluons. Soft
hadrons from emitted gluons become significant only at small $z_T$.
At large $z_T$ hadrons mainly come from fragmentation of the jet
with reduced energy. Thus if one chooses large $z_T$, the near-side
jet profile should not change. On the other hand, the back-side
profile could change due to transverse momentum broadening.

\section{High $p_T$ azimuthal anisotropy}

In non-central heavy-ion collisions, the parton energy loss
has finite azimuthal anisotropy due to the azimuthal
dependence of the path length of propagation. This will
lead to a large azimuthal anisotropy of high-$p_T$ hadron
spectra \cite{wangv2} which has been observed by RHIC
experiments \cite{star-v2}. After
correction for two-particle correlations, the observed
azimuthal anisotropy is consistent with that caused by
parton energy loss \cite{wang03}. The same energy loss also
explains quantitatively the single hadron suppression and 
suppression of back-side jet correlations.

Since azimuthal anisotropy in hadron spectra is generated by
the geometrical eccentricity of the dense medium, it is only
sensitive to the evolution of the dense matter at very early
time \cite{olli}. As the system expands, the geometry becomes 
more symmetric and thus loses its ability to generate 
spectra anisotropy. This is particularly true in late hadronic
stage \cite{kolb}. If there were no parton energy loss and 
no jet attenuation before a finite hadron formation time, then
any anisotropy in spectra will be caused by the geometrical
eccentricity at the time when hadron absorption starts.
At this late time, a few fm/$c$ for example, the geometry
is already quite symmetric and can no longer generate
large anisotropy in the final hadron spectra.
Therefore, the observed large azimuthal 
anisotropy at high $p_T$ cannot be generated by hadronic
absorption of jets in the late stage of the evolution.

\section{SPS data}

\begin{figure}
\centerline{\psfig{figure=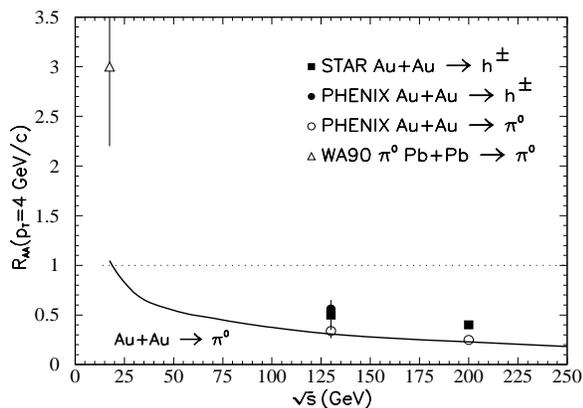,width=3.0in,height=2.1in}}
\caption{The colliding energy dependence of the nuclear modification
factor for single inclusive hadron spectra at fixed $p_T$ in the
most central $Au+Au$ (or $Pb+Pb$) collisions as compared
to the parton model calculation. The parton energy loss
is assumed to be proportional to the measured charge 
multiplicity $dN_ch/d\eta$ while the medium formation 
time and lifetime of the medium are assumed to be the same. 
The data are from PHENIX \protect\cite{phenix-r}, STAR \protect\cite{star-r} 
and WA98 \protect\cite{wa98}.}
\label{fig7}
\end{figure}

The final piece of the evidence comes from experiments at SPS. 
Hadron spectra at this energy are very steep at high $p_T$ and
are very sensitive to initial transverse momentum broadening
and parton energy loss \cite{wang98}. However, the measured
$\pi$ spectra in central $Pb+Pb$ collisions only show the 
expected Cronin enhancement \cite{wa98,wang98h} with no sign 
of significant suppression. More recent analyses of the $Pb+Pb$
data at the SPS energy also show \cite{ceres} that both same-side 
and back-side jet-like correlations are not suppressed, though the
back-side distribution is broadened. Shown in Fig.~\ref{fig7}
as a solid line is the energy dependence of the calculated
single pion suppression factor at $p_T=4$ GeV/$c$ as compared to 
data at RHIC and SPS. The initial gluon density
$\rho_0$ at $\tau_0=0.2$ fm/$c$ in the calculation of the parton 
energy loss in Eq.~(\ref{eq:de}) is assumed to be proportional 
to the measured $dN_{\rm ch}/d\eta$ \cite{phobos-s}. The measured
multiplicity at SPS is only about 2.0 smaller than at the 
highest energy of RHIC. The calculated suppression 
increases more rapidly and reaches at 1 at the SPS energy.
This is partly because of the Cronin effect which is much stronger
at SPS and compensates some of the energy loss effect. However, the
calculation is still about a factor of 3 smaller than the data.
Similar results are reported in Ref.~\cite{gv02} when the same
gluon density is used.

There could be several reasons for such a big discrepancy between 
data and our calculation at SPS \cite{wang98}. The initial formation
time $\tau_0$ could be much larger than at RHIC or the lifetime of 
the dense matter at SPS could be much shorter. A critical behavior
of the screening mass, which influence parton energy loss, could
also lead to a sudden decrease of energy loss \cite{dumitru}
at lower energies. Since a hadronic 
gas should have at least existed in $Pb+Pb$ collisions at SPS and 
the particle 
density and duration of such a hadronic state should not be much
different from that in $Au+Au$ collisions at RHIC, hadronic
rescattering or absorption should have significantly suppressed
the pion spectra, were it responsible for most of the jet
quenching at RHIC. Therefore, in any circumstances, the SPS
data are not consistent with a hadronic absorption picture at RHIC.

Nevertheless, jet quenching at SPS energies still remains a less
explored territory. As shown in Fig.~\ref{fig7}, it will be 
important to have a few measurements between SPS and RHIC energy
to explore the colliding energy dependence of jet quenching and
find out whether there is any threshold behavior of jet quenching.
By changing the colliding energy, one essentially changes the
the initial parton density without changing the initial medium
size. This will allow one to observe the initial density dependence
of jet quenching, obtain more information about formation time
or lifetime of the medium, and search for critical behaviors that
might be caused by phase transitions in the evolution of the
dense medium.

\section{Summary}

This paper has provided arguments and listed experimental evidence
that the observed jet quenching in $Au+Au$ collisions at RHIC
is caused mainly by parton energy loss, not hadron absorption
in a hadronic medium. The estimated hadron formation time in
jet hadronization is much longer than the typical lifetime of
the dense matter and thus cannot be responsible for the
observed jet quenching.
The observed $p_T$ and centrality dependence of jet quenching are
not consistent with a hadronic absorption picture with a finite formation
time that is smaller than the size of the medium. The measured
high-$p_T$ azimuthal anisotropy can only be caused by the geometrical
anisotropy of the medium in a very early stage and thus cannot be
due to hadronic rescattering. The most direct evidence for partonic
energy loss and jet hadronization outside the medium is the universal
same-side leading hadron correlations inside a jet in $p+p$, $d+Au$
and $Au+Au$ collisions. Hadronic rescattering or absorption inside
the medium would have destroyed the jet-like same-side correlation. 
Finally, the absence of jet quenching in $Pb+Pb$ collisions at SPS also
indirectly proves that hadronic rescattering cannot be responsible
for the observed jet quenching at RHIC.

A direct measurement of parton energy loss is proposed which requires
the reconstruction of the total energy of a jet that has a triggered 
hadron with
a fixed value of $p_T^{\rm trig}$. The difference between $Au+Au$
and $p+p$ measurements (plus $p_T$ broadening due to initial
multiple parton scattering) should be related to the averaged total
energy loss for the jet whose leading parton produces the triggered
hadron after energy loss. The measurement of softening of 
the effective hadron-triggered fragmentation function will further 
detail the pattern of energy loss and induced gluon emission. 
The importance of jet quenching studies at lower RHIC energies 
is also discussed.

\section*{Acknowledgement}

Some ideas and arguments presented here have been discussed by many people
in talks and informal discussions which cannot be tracked back
exactly. The author has benefitted, in particular, from discussions with
D. Hardtke, P. Jacobs, H.-G. Ritter, V. Koch, Y. Kochevgov and 
B.~Z.~Kopeliovich. The author is indebted to F. Wang for
discussions that lead to the idea of direct measurement of 
energy loss.
This work was supported by the Director, Office of Energy
Research, Office of High Energy and Nuclear Physics, Divisions of 
Nuclear Physics, of the U.S. Department of Energy under Contract No.\
DE-AC03-76SF00098.

\end{multicols}

\end{document}